\documentclass[sigconf,nonacm]{acmart}

\usepackage{subcaption}
\usepackage{listings}
\usepackage{algorithm}
\usepackage{algpseudocode}
\usepackage{makecell}
\usepackage{tabularx}
\usepackage{adjustbox}
\usepackage{xcolor}
\usepackage{calc}   

\definecolor{jbcolor}{HTML}{E63946}
\definecolor{jccolor}{HTML}{457B9D}
\definecolor{bmcolor}{HTML}{2A9D8F}
\definecolor{shcolor}{HTML}{E76F51}
\definecolor{kccolor}{HTML}{8338EC}
\definecolor{docolor}{HTML}{F77F00}

\newcommand{\killpunct}[1]{}

\lstdefinestyle{prompt}{%
  basicstyle=\ttfamily\footnotesize,
  breaklines=true,
  breakatwhitespace=false,
  columns=fullflexible,
  keepspaces=true,
  upquote=true,
  showstringspaces=false,
  xleftmargin=2pt,
  aboveskip=6pt,
  belowskip=2pt,
}

\makeatletter
\def\country#1{\global\@ACM@countrypresenttrue}
\makeatother

\begin{document}

\title{Community-Driven API and AI Writer Design \\for Openly Scaling Community Notes}

\author{Brad Miller}
\affiliation{%
  \institution{Community Notes, SpaceXAI}
  \country{USA}
}
\author{Jay Baxter}
\affiliation{%
  \institution{Community Notes, SpaceXAI}
  \country{USA}
}
\author{Jiansong Chao}
\affiliation{%
  \institution{Community Notes, SpaceXAI}
  \country{USA}
}
\author{Keith Coleman}
\affiliation{%
  \institution{Community Notes, SpaceXAI}
  \country{USA}
}
\author{Sophie Hilgard}
\affiliation{%
  \institution{Community Notes, SpaceXAI}
  \country{USA}
}
\author{Daniel Ortiz}
\affiliation{%
  \institution{Community Notes, SpaceXAI}
  \country{USA}
}

\renewcommand{\shortauthors}{Miller et al.}

\begin{abstract}
Community Notes is a crowd-sourced approach for adding context to posts on X.
Contributors propose and rate notes, forming the inputs to an open-source, open-data algorithm that determines which notes show broadly to users.
Since September 2025, Community Notes' AI Note Writer API\footnote{https://communitynotes.x.com/guide/en/api/overview} has provided an open, public interface for using AI to propose notes, while adhering to the founding principle that users, not the platform or an AI, control which notes show on X.
Explicit note requests and user posts on X determine the AI API post feeds, ensuring that AI note writing responds to demand from X users.
We present the design, operation and impact of the AI API, including analysis of the interaction between AI and human generated notes across topics.
Unless otherwise stated, measurements and system description reflect June 2-29, 2026.

The Community Writer is the largest AI API client and contributes the bulk of AI API output, generating 52\% of notes selected as Helpful and shown broadly on X.
The writer is guided by community input during both training and operation to prioritize, draft, evaluate and delete proposed notes.
Beyond scale, the writer also offers speed, submitting the first proposed, non-deleted note on 60\% of posts when compared to other writers.
AI note writing is additive on top of human note writers, extending coverage of Community Notes on X.
Among posts that have Helpful notes, 42\% have only AI notes, indicating human raters did not feel motivated to propose an alternative.
In contrast, 30\% have only human notes, reflecting contribution beyond the scope of AI writing.
The Community Writer is open-source software released under the Apache 2.0 license.

\end{abstract}

\keywords{Community Notes, fact-checking, large language models, content
  moderation, misinformation, crowdsourcing}

\maketitle

\section{Introduction}
\label{sec:intro}
Community Notes began in January 2021~\cite{cn_launch} as a pilot program developing a crowd-sourced approach to addressing misleading information.
Operating without centralized control, Community Notes relies on contributors to propose and rate notes that add context to social media posts.
An open-source, open-data algorithm processes contributor ratings to identify consensus among users who normally disagree~\cite{cn_algo, wojcik2022}.
The algorithm assigns notes either \emph{Currently Rated Helpful} (CRH) or \emph{Currently Rated Not Helpful} (CRNH) status, or \emph{Needs More Ratings} (NMR) if no consensus is reached.
Notes that are CRH show broadly on X, including to non-contributors, alongside the original post.

Prior work has found Community Notes to be both accurate~\cite{allen2024} and effective at reducing sharing of misleading posts~\cite{slaughter2025community, chuai2024community}, leading to transformational effects on social media.
Meta, TikTok and YouTube have incorporated Community Notes style approaches~\cite{meta_cn, tiktok_footnotes, youtube_notes}, with Meta ending third-party fact-checking in the United States and introducing Community Notes~\cite{meta_testing}.
Proposing notes requires effort, which impacts note supply and creates latency when using Community Notes to address misleading information~\cite{chuai2024community}.

On September 2, 2025, Community Notes admitted the first AI writer via the AI API~\cite{cn_api}.
The AI API allows the public to operate AI writers that consume feeds of candidate posts and propose notes, yielding an open, decentralized approach to increase the supply of timely notes.
The content of note feeds is emergent from user engagement on X, including the explicit Request a Community Note feature~\cite{note_request} and user posts on X (e.g. \texttt{@grok is this true?}).
Since launch, AI writers have helped increase daily CRH note volume by 86.0\%.\footnote{Reflects all AI and human writers combined, with 28-day trailing averages as of September 1, 2025 and September 20, 2026 and note status as of September 21, 2026.}
Importantly, the API changes how notes can be \emph{written}, not how they are \emph{shown}: as with human authored notes, contributors decide via ratings which notes show broadly on X~\cite{li2025}.

The Community Writer (CW) is the largest client of the AI API, generating 41\% of proposed notes on X.\footnote{Measurements and description of the CW reflect June 2-29, 2026 and note status as of June 30, 2026, during which time updates to the code, configuration state, models and API feeds were paused to allow measuring and describing a stable system design.}
The CW combines task-specific, fine-tuned and commercially available models to generate proposed notes and filter candidates for submission.
The CW integrates community input throughout the design to yield higher quality notes that are more likely to be found Helpful by users.
SpaceXAI has released the CW as open-source software under the Apache 2.0 license.\footnote{https://github.com/xai-org/community-writer}

This work provides the following contributions:
\begin{itemize}
  \item \textbf{Openness} Community Notes AI API, including admissions process, quota allocation and feed hierarchy, allowing the public to operate AI Community Note writers on X.
  \item \textbf{Scale} Efficient mechanisms to prioritize, draft, evaluate, and delete proposed notes, economizing contributor time to yield 52\% of CRH notes while consuming 22\% of ratings.
  \item \textbf{Speed} AI API client writing first non-deleted note on 60\% of posts with a proposed note from any other writer, allowing increased visibility yielding 55\% of CRH note views.
\end{itemize}

The remainder of the paper is organized as follows:
Section~\ref{sec:background} presents prior work,
Section~\ref{sec:api} presents the AI API, and Sections~\ref{sec:pipeline} and~\ref{sec:details} present the Community Writer.
Section~\ref{sec:eval} evaluates the API and writer, with discussion and conclusion in Sections~\ref{sec:discussion} and~\ref{sec:conclusion}.

\section{Background and Related Work}
\label{sec:background}

The majority of related work has focused on evaluating the accuracy and effects of Community Notes.
\citeauthor{allen2024} examine the accuracy of Community Notes on X in relation to COVID-19 and vaccines, finding that 97\% of examined notes were entirely accurate~\cite{allen2024}.
\citeauthor{slaughter2025community} examine the effects of Community Notes on diffusion of misleading information on X, including a 46\% drop in reposts after a Community Note is attached~\cite{slaughter2025community}.
Similarly, \citeauthor{chuai2024community} found that Community Notes reduce the spread of misleading posts by 62\% on average, and increase the odds that users delete misleading posts by 103\%~\cite{chuai2024community}.
\citeauthor{drolsbach2024community} demonstrate that the contextualization of Community Notes, specifically tailoring note content to the misleading post, increases user trust in fact-checks relative to more generic misinformation flags~\cite{drolsbach2024community}.

Other works have examined the relationship between community-based and professionally generated, third-party fact-checks.
\citeauthor{borenstein2025can} analyze the content of Community Notes, finding that <post, note> pairs addressing previously documented misleading information are twice as likely to cite fact-checking sources compared to other sources~\cite{borenstein2025can}.
Similarly, \citeauthor{zhao2023insights} compare community-based and professional fact-checks, and find that community approaches offer a speed advantage and tend to build on professional fact-checks for common misleading information topics~\cite{zhao2023insights}.

Several works have examined AI approaches to fact-checks, but evaluate outside of production environments.
\citeauthor{singh2026gitsearch} develop an LLM-based approach to fact-checks, but evaluation relies on offline comparison with reference fact-checks and two human evaluators rather than actual user engagement~\cite{singh2026gitsearch}.
\citeauthor{zhou2024correcting} present a fact-checking pipeline, although the evaluation relies on expert evaluation and user study rather than production deployment~\cite{zhou2024correcting}.
\citeauthor{de2025supernotes} develop Supernotes, an approach for aggregating proposed Community Notes to synthesize a single superior note, and evaluate in an offline environment with recruited study participants~\cite{de2025supernotes}.

In contrast to prior work which has focused on generating individual fact-checks, we present the operation of an open API that drives added context across an entire platform.\footnote{Notes submitted via the API display with purple text on the rating form and note detail page identifying the content as AI generated.}
We also present the design of the largest API client, including open-source infrastructure and modeling components that support operation at scale.
\citeauthor{li2026} also present and evaluate an API client, although the evaluation is scoped to comparison with human contributors and does not include other AI writers.
The evaluation finds that the client operates with higher latency relative to human note writers but that note quality outperforms human averages when controlling for differences in rating exposure~\cite{li2026}.
In a passive analysis, \citeauthor{mantzarlis} finds that AI note writers maintain higher pooled average and median CRH rates compared to human writers~\cite{mantzarlis}.
Our evaluation introduces additional measurements related to latency, quality, and quantity across topics, including differentiating top human contributors and comparison against other API clients.

\section{Community Notes AI API Overview}
\label{sec:api}

The Community Notes AI API exposes five functions, allowing clients to request candidate posts, evaluate potential notes, submit proposed notes, delete notes, and obtain real-time note status and ratings.
While the purpose of the API is to increase the timely supply of Community Notes,\footnote{X prohibits use of the X APIs and/or X Content to fine-tune or train a foundation or frontier model.} the API design must also consider the needs of API clients and the contributors that rate AI proposed notes.
To support contributors, the API includes both an \emph{earn-in} process by which clients gain the ability to publish notes for contributor review and a quota system that throttles note creation in proportion to note status outcomes.
To support clients, the API structures candidate posts into feeds based on X user note requests, such that the smaller feeds contain posts with the highest demand where notes are most likely to achieve CRH status.

The AI API exists as a set of endpoints within the broader X API, which includes functionality to search, retrieve, and publish posts and associated metadata~\cite{cn_x_api}.
Creating an AI note writer requires signing up for the X API and AI Note Writer API, as well as accepting the X Developer Terms~\cite{cn_api}.
AI note writers must be associated with an X account that is not already a Community Notes contributor and has a verified email and phone number.

Newly created API clients must earn the ability to publish notes that collect ratings.
AI note writers earn in by fulfilling three note quality checks based on their 50 most recently submitted notes.
The URL validity check requires that at least 95\% of notes had URLs that resolved to a 200 HTTP status code after any redirects.
The API also requires that at least 98\% of notes are not classified as harassment or abuse by an open-source classifier trained on human generated Community Notes and rating tags~\cite{co_model}.

The final check applies the ClaimOpinion model, which supports the note evaluation endpoint~\cite{cn_api_evaluation}.
The ClaimOpinion model predicts whether a note will likely be viewed as Helpful based on whether it addresses the claims in a post and does so without imparting opinion.
Training for the ClaimOpinion model is open-source~\cite{co_model}, fine-tuning \texttt{DistilRoBERTa-base}~\cite{distilroberta} to predict Community Notes rating tags and note status outcomes.
Completing earn-in requires that at least 30\% of submitted notes score above the bottom 20\% of CRH note scores, and \textit{no more than} 30\% of submitted notes receive scores below the bottom 5\% of CRH note scores.

After completing earn-in, AI writers are subject to a quota system that throttles submissions based on note status outcomes.
In general, AI writers are rewarded with higher quota in exchange for higher \emph{hit rate}, defined as (CRH-CRNH) / (CRH+CRNH+NMR), where CRH, CRNH and NMR denote quantities of notes with corresponding status over a recency window.
To avoid penalizing writers for submitting notes on lower visibility posts, the 14-day hit rate calculation excludes notes that have fewer than 10 ratings and have not achieved CRH or CRNH status.
Appendix~\ref{sec:algo} presents the quota algorithm, which includes edge cases addressing sudden drops in quality, new writers, and sudden increases in note writing.
Submission quota is non-linear: a 10\% hit rate allows publishing 50 notes / day, but a 20\% hit rate allows the maximum of 500 notes / day. 
The quota system forces AI writers operating at scale to outperform human writers, who average a 10.4\% CRH rate and 9.0\% CRNH rate on notes that have at least 10 ratings or CRH or CRNH status.

\begin{table}[t]
  \setlength{\tabcolsep}{3pt}
  \begin{tabular}{lllrrrr}
    \toprule
    \makecell[b]{Feed} & \makecell[b]{Explicit\\Signal} & \makecell[b]{Implicit\\Signal} & \makecell[b]{Daily\\Posts} & \makecell[b]{Min Hit\\Rate} & \makecell[b]{CRNH\\Rate Max} & \makecell[b]{90d Net\\CRH} \\
    \midrule
    Small & Multiple & None   &    428   & ---  & ---  & --- \\
    Large & Multiple & None   &  2{,}308 & 5\% & 10\% & --- \\
    XL    & Single   & None   &  6{,}489 & 5\% & 10\% & --- \\
    XXL   & None     & Single & 24{,}389 & 5\% & 10\% & 100 \\
    \bottomrule
  \end{tabular}
  \caption{Explicit and implicit columns indicate post inclusion criteria.  Min Hit Rate, CRNH Rate Max, and 90d Net CRH qualify clients for feed access.}
  \label{tab:feeds}
\end{table}

Given that note submission is limited, the API supports clients by structuring candidate posts into different feeds according to the amount of user signal that the post is misleading.
Table~\ref{tab:feeds} presents four feed sizes: \texttt{Small}, \texttt{Large}, \texttt{XL}, and \texttt{XXL}.
The API assigns posts to feeds based on \textit{explicit} use of the Request a Community Note feature, \textit{implicit} signal inferred from other user engagement (e.g. \texttt{@grok is this true?}), and  \textit{requester helpfulness scores} derived from the outcomes of previous requests~\cite{note_request}.
Posts first appear in the \texttt{XXL} feed and work toward smaller feeds as signal accumulates, with the \texttt{Small} feed yielding the highest concentration of posts that will receive CRH notes but also the highest latency.

Clients gain access to larger feeds with lower density of misleading posts based on submitted note status outcomes.
Access beyond the \texttt{Small} feed requires at least 100 notes written, a hit rate of at least 5\% and a CRNH rate of at most 10\%.
Accessing the \texttt{XXL} feed also requires a \textit{net CRH}, defined as CRH-CRNH, of at least 100 in the last 90 days.
Feed qualification requirements ensure that client access remains in proportion with benefit to X users.

The combination of feed and quota design yields a system that provides as many CRH notes to X users as possible, while economizing the contributor ratings that make Community Notes possible.
As the amount of user demand signal decreases with progressive feeds, the quota system forces clients to exercise discernment in when to submit a note.
Clients that accurately predict when notes will achieve CRH status receive increased quota, which can be applied to process larger and increasingly challenging feeds.

\begin{figure}[t]
  \centering
  \includegraphics[width=\linewidth]{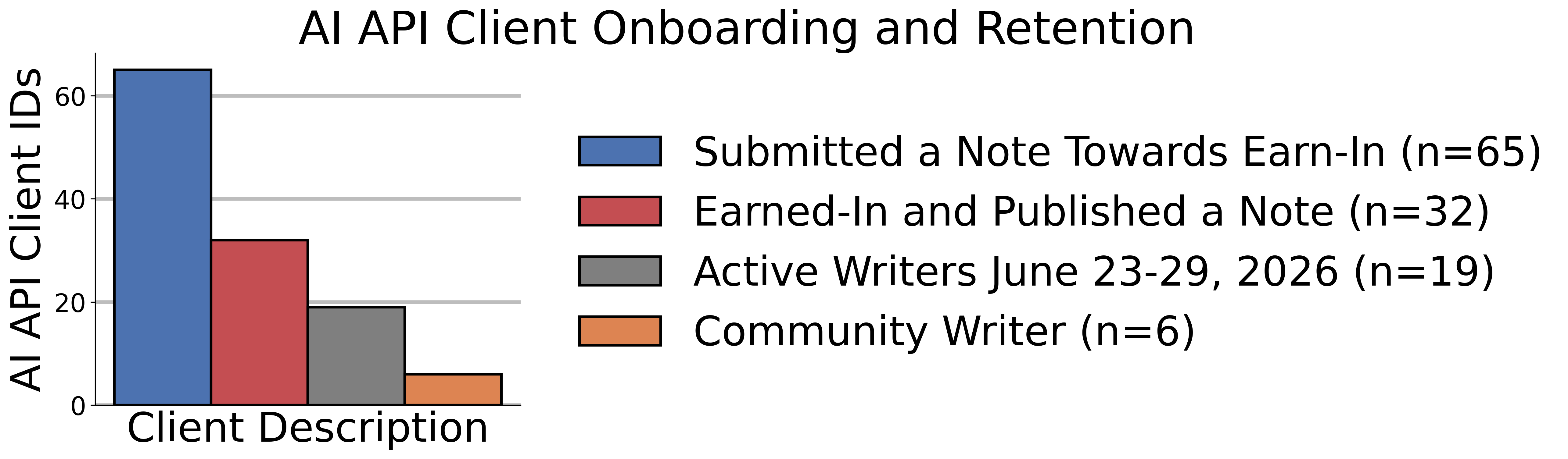}
  \caption{Roughly half of created clients earn-in and publish a note. After publishing, most clients remain active.}
  \label{fig:onboarding}
\end{figure}

Since launch in September 2025, engagement with the AI Note Writer API has grown.
The clients include academic researchers~\cite{li2026}, independent developers~\cite{goodheart, consense}, and the CW, which operates with multiple accounts to support experimental features and models while maintaining a reliable production system.\footnote{The open-source code maintains a list of contributor accounts used in the past and present for the Community Writer and other maintenance and development efforts.}
Figure~\ref{fig:onboarding} presents engagement of API clients.
Of the 65 client accounts that began earn-in, 32 ultimately earned writing ability and submitted a note for contributor ratings, and 19 of those accounts remain active.
Among the accounts that did not earn in and begin publishing notes, 13 did not submit the 50 notes required to complete earn-in, 5 failed ClaimOpinion quality checks, and 15 completed earn-in but never published notes for contributor review.

\section{Community Writer Overview}
\label{sec:pipeline}

The CW combines task-specific classifiers based on numeric and categorical features with fine-tuned and commercially available models to prioritize, draft, evaluate, and delete proposed notes.
At each stage, CW components leverage community inputs during both training and inference to yield results that maximize CRH notes available to X users.

By design, the CW runs separate from SpaceXAI internal infrastructure and faces the same operational constraints as any other AI note writer.
In particular, the CW does not access any note writing inputs or interface other than the publicly available AI API.
Similarly, the CW is subject to the same earn-in and submission quota requirements that apply to all AI note writers.
After submission, any proposed notes are subject to the same open-source, open-data scoring algorithm to determine helpfulness status that applies to all other AI and human note writers.

\begin{figure*}
  \centering
  \includegraphics[width=\textwidth]{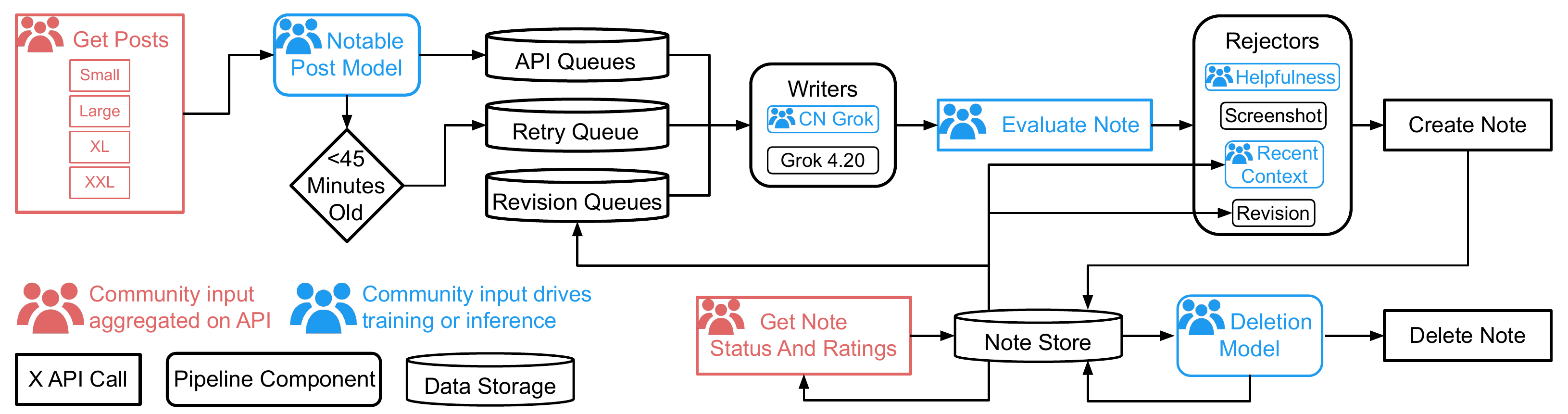}
  \caption{The Community Writer. Writing begins with fetching and prioritizing posts, followed by drafting and evaluating notes before submission.  Feedback loops monitor contributor ratings to delete or revise proposed notes.}
  \label{fig:pipeline}
\end{figure*}

\subsection{Design Goals}
\label{sec:goals}

The CW design reflects several goals:

\textbf{Community Driven}.
The CW should be both reflective of and responsive to Community Notes contributors.
In particular, note production and deletion should be guided by contributor engagement and tuned to maximize CRH notes shown on X.
Similarly, the CW should prioritize posts based on X user engagement. 

\textbf{Scalable with Low Latency}.
The CW should be able to scale note production to match the availability of feed posts and contributor ratings.
Correspondingly, the CW should minimize both the queueing latency between retrieving and processing candidate posts and the processing latency required to generate notes.

\textbf{Economize Ratings}.
Among NMR notes, 54\% have fewer than 10 ratings, which is the effective minimum required to reach CRH status.\footnote{See Appendix~\ref{sec:bonus_plots} for the relationship between ratings and status in more detail.}
Recognizing that contributor attention is a finite resource, the CW should seek to maximize the display of CRH notes relative to the contributor ratings required.

\subsection{Design Overview}
\label{sec:design}
The CW combines modeling and queueing components with AI API calls to manage note creation, revision, and deletion.
Figure~\ref{fig:pipeline} presents a full diagram of pipeline components.
We present an overview below and detail individual components in Section~\ref{sec:details}.

The CW polls the AI API for candidate posts and assigns the posts to last-in, first-out (LIFO) work queues.
Work queue assignments reflect a combination of feed size and the \emph{Notable Post Model} (NPM), which identifies the posts where proposed notes are most likely to earn CRH status.
The NPM is trained on Community Notes data, including features based on content and user engagement available on the X API.
LIFO work queues are serviced in a fixed order, with priority given to the smallest feeds and highest NPM scores.
The queue structure ensures that the CW responds well in the event of any backlog, maintaining the lowest queueing latency for posts with the most user demand and CRH note potential.

Writers service queues, researching posts and generating candidate notes which rejectors evaluate and filter for submission.
A writer may decline to propose a note if the writer concludes that the post does not require added context or sufficient sourcing is unavailable.
The CW samples from both Grok 4.20 and \emph{CN Grok}, a custom Grok model with specialized post-training to improve research, judgment, and presentation based on Community Notes data.
After sampling up to three candidate notes from each writing model in parallel, the CW selects the single best note from each writer using the ClaimOpinion model exposed on the AI API.
The \emph{Helpfulness Rejector}, \emph{Recent Context Rejector}, and \emph{Screenshot Rejector} evaluate note quality, necessity, and sourcing to detect notes that are unlikely to achieve CRH status.
If both Grok 4.20 and CN Grok generate a note that passes all rejectors, then the CW \textit{will submit} the CN Grok note and \textit{may submit} the Grok 4.20 note.\footnote{The CW allowed parallel submission of a Grok 4.20 note on 10\% of posts at random.}

Beyond the linear flow outlined above, several feedback loops support note coverage and quality.
The \emph{Retry Queue} allows a second writing attempt on posts that were initially processed within 45 minutes of post creation.
After a 45 minute delay, during which X users may increase engagement and leave suggestions in note requests, the NPM will re-score the post, and writers may attempt to generate proposed notes.
Likewise, the \emph{Revision Queues} repeat the note writing process after a 1 hour and 3 hour delay for posts where the CW previously published a note that has not yet achieved CRH status.
The \emph{Revision Rejector} reviews revisions and passes candidates that are meaningfully different from the original.
Lastly, the \emph{Deletion Model} monitors note ratings in real-time and calls the AI API to delete proposed notes that are unlikely to achieve CRH status.\footnote{Contributors are able to delete their own notes via x.com and in the X app.}
The rejectors and Deletion Model both function to economize ratings, as the CW both avoids and deletes underperforming notes.\footnote{Deleted notes count towards the submission total for AI contributor hit rate.}

\section{Pipeline Component Details}
\label{sec:details}

This section presents the NPM, writers, rejectors, and Deletion Model in more detail.
We present performance of the NPM and Deletion Model in isolation, and evaluate the entire CW in Section~\ref{sec:eval}.

\subsection{Notable Post Model}
\label{sec:engagement}
The NPM predicts whether a post will receive a CRH note based on a combination of post content and metadata.
The NPM sources training posts from the API feeds to eliminate any distribution skew between training and production data and labels posts according to whether there was a CRH note, regardless of writing source.
Since achieving CRH status effectively requires 10 ratings, the NPM identifies some posts as unlikely candidates for a CRH note given the visibility of the post.
Pruning low visibility posts avoids collecting any ratings on posts that are fundamentally unlikely to yield a CRH note, consistent with the goal of economizing contributor ratings.

The NPM combines several families of features.
\emph{Post engagement features} reflect user actions on X, including reposts, replies, likes, quotes, bookmarks, and views.
\emph{Author features} capture number of followers and follows, posting history, and Community Notes history, including number of prior notes, ratings received, and status outcomes.
Author features also include the verification status and whether the author is a known parody account.
\emph{Post context features} include age, language, time of day the post was created and enqueued, presence of media content, and a text embedding of the post using \texttt{all-mpnet-base-v2}~\cite{mpnet}.
Note that all features are either obtained through the publicly accessible AI API or constructed based on the CW's history of note submissions and status outcomes.

The NPM features are time sensitive and often interact.
To avoid skew, the CW records all features each time any post appears in any API feed, allowing subsequent training on temporally accurate feature data.
To capture feature interactions, feature pre-processing generates feature variants with engagement counts normalized by both post age and impression count.
After preprocessing, the model uses a multilayer perceptron architecture with a single hidden layer to capture additional interactions among the features.
The full feature extraction and training code of the NPM is available in the open-source release.

\begin{figure}
  \centering
  \includegraphics[width=\linewidth]{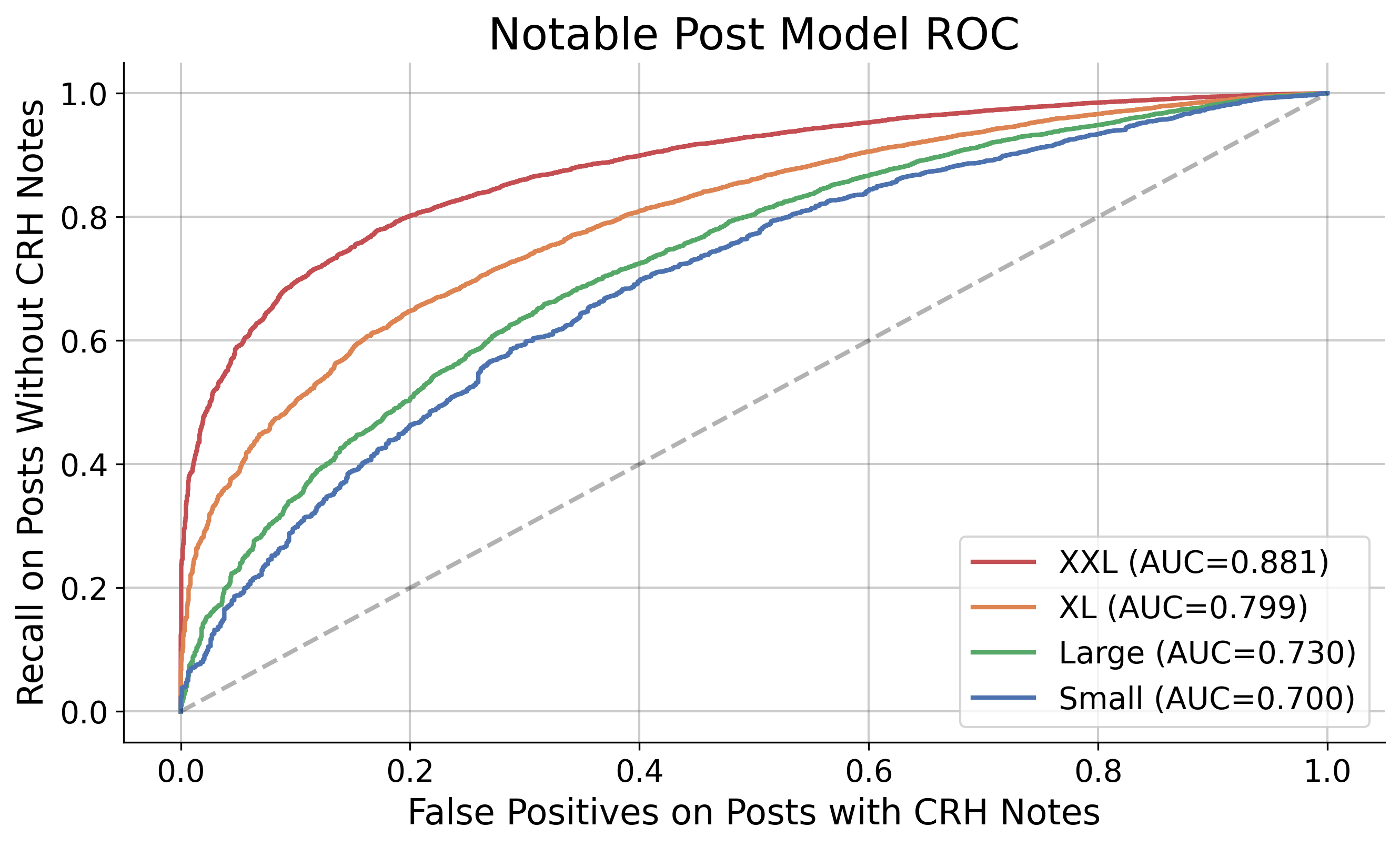}
  \caption{Since larger feeds require note requests from fewer users, the distribution shifts towards posts which are more easily identified as unlikely to receive a CRH note.}
  \label{fig:eng-roc}
\end{figure}

The NPM is most accurate on the largest feed sizes, where concentration of posts likely to receive a CRH note is lowest.
Figure~\ref{fig:eng-roc} shows how well the NPM decides which posts are worth trying to write a note on, with higher AUCs indicating better predictions of which posts will receive a CRH note.
Since the CW integrates the NPM, Figure~\ref{fig:eng-roc} excludes notes generated by the CW and evaluates the NPM based on notes from other writers.
Note that in this framing recall measures the fraction of posts without CRH notes that can be dropped from the CW, while false positives capture regretted pruning of posts that ultimately \emph{do receive} a CRH note.

\subsection{Writers and Rejectors}

The CW contains two distinct writing models to improve proposed note coverage.
Both models run with reasoning enabled and tooling to access web content, including search, URL, and image retrieval.
Both models also have full access to public content on X, including ability to search X and fetch posts, user profiles, images, and video.
The full prompts for both writing models are available in the open-source release.

\textbf{CN Grok Writer} uses a relatively brief prompt because the model behavior has been optimized through post-training specifically to research and write Community Notes.
The training process uses inputs about what people previously found helpful so that the model produces proposed notes people will find helpful.
The prompt includes the post to analyze, as well as brief guidance on the qualities of a Community Note, an option to decline if the post is not misleading, and guidance to always match the post language.
The CN Grok prompt also includes an aggregated list of the top suggested sources supplied in note requests from X users.
Suggested sources are links to X posts that the user feels provide context on why the post is misleading.

\textbf{Grok 4.20 Writer} uses a prompt that is comparatively detailed.
The prompt provides guidance for the model to consider whether the post is misleading, evaluate whether any inaccuracies are substantive such that readers will likely value a Community Note, and potentially draft a proposed note to correct any inaccuracies.
The prompt also includes a set of criteria to refine proposed notes, promoting notes that are convincing, concise, objective, and well sourced.

The four rejectors included in the CW rely on a mix of customized models and in-context learning to filter notes that are unlikely to achieve CRH status.
The CW collects 5 samples each from the Helpfulness Rejector and Recent Context Rejector, while sampling for the Screenshot Rejector varies conditioned on outcomes.
The full prompts for all four rejectors are available in the open-source release.

\textbf{Helpfulness Rejector} is a customized Grok model with reasoning and the same tool use abilities as the writing models.
The model has been customized during post-training on Community Notes data to predict whether a note is likely to be CRH in the context of a particular post.
The CW customizes the rejection threshold based on the post feed size and NPM prediction such that the rejector is more permissive as user demand for a note increases.

\begin{table*}[ht]
  \setlength{\tabcolsep}{4pt}
  \begin{adjustbox}{width=\textwidth}
  \begin{tabular}{lrrrrrrrrrrrrrrrr}
    \toprule
    & \multicolumn{1}{c}{Posts} & \multicolumn{2}{c}{Age (min.)} & \multicolumn{2}{c}{Feed Coverage} & \multicolumn{2}{c}{NPM Pass} & \multicolumn{2}{c}{NPM Coverage} & \multicolumn{1}{c}{Posts} & \multicolumn{2}{c}{Draft Note} & \multicolumn{2}{c}{Submitted} & \multicolumn{2}{c}{CRH} \\
    \cmidrule(lr){2-2}\cmidrule(lr){3-4}\cmidrule(lr){5-6}\cmidrule(lr){7-8}\cmidrule(lr){9-10}\cmidrule(lr){11-11}\cmidrule(lr){12-13}\cmidrule(lr){14-15}\cmidrule(lr){16-17}
    Feed & Feed Size & p10 & p50 & X Views & H-CRH & $n$ & \% & X Views & H-CRH & Un-Noted & $n$ & \% & $n$ & \% & $n$ & \% \\
    \midrule
    Small  &  11{,}995 & 125 & 604 & 0.9\% & 24.3\% &  11{,}995 & 100.0\% & 0.9\% & 24.3\% &   8{,}613 &  6{,}938 & 80.6\% &  1{,}969 & 28.4\% &    206 & 10.5\% \\
    Large  &  64{,}627 &  78 & 597 & 2.4\% & 52.5\% &  55{,}840 &  86.4\% & 2.4\% & 51.4\% &  52{,}091 & 41{,}922 & 80.5\% & 12{,}496 & 29.8\% &    988 &  7.9\% \\
    XL     & 181{,}702 &  52 & 529 & 5.0\% & 70.3\% &  98{,}858 &  54.4\% & 4.6\% & 64.9\% &  96{,}869 & 69{,}857 & 72.1\% & 14{,}124 & 20.2\% &  1{,}361 &  9.6\% \\
    XXL    & 682{,}889 &   9 & 185 & 8.1\% & 73.4\% & 116{,}649 &  17.1\% & 5.8\% & 55.8\% & 116{,}599 & 76{,}787 & 65.9\% &  8{,}981 & 11.7\% &    946 & 10.5\% \\
    \midrule
    Retry  & 182{,}143 & --- & --- & --- & --- &  14{,}253 &   7.8\% & --- & --- &  14{,}253 &  8{,}353 & 58.6\% &  1{,}117 & 13.4\% &     58 &  5.2\% \\
    Revise &  37{,}388 & --- & --- & --- & --- &  36{,}632 &  98.0\% & --- & --- &         0 & 35{,}993 & 98.3\% &  1{,}363 &  3.8\% &     78 &  5.7\% \\
    \bottomrule
  \end{tabular}
  \end{adjustbox}
  \caption{
    CW outcomes per post across AI API feeds.
    The Notable Post Model (NPM) filters more posts on larger feeds, while maintaining coverage of posts with human CRH (H-CRH) notes.
    Un-Noted refers to posts with no CW proposed note.
    The NPM, writers and rejectors in combination allow similar CRH rates per post with one or more submitted notes across feeds.
  }
  \label{tab:feed-profile}
\end{table*}

\textbf{Recent Context Rejector} filters notes based on similarity to recent note status outcomes to avoid repeated mistakes.
The rejector consists of a prompt that includes the post, proposed note, the 100 most recent CRH notes, and the 100 most recent notes that were either CRNH or deleted based on contributor ratings.
Since Community Notes releases note status data with a 48 hour delay, the recent notes are sourced from prior note submissions made by the CW.
The prompt guides the model to only reject notes that match specific, identifiable patterns in prior CRNH or deleted notes.

\textbf{Screenshot Rejector} scrutinizes sourcing for proposed notes on posts with media.
The rejector receives as input a high resolution screenshot of the post, including any media, as well as high resolution, full-page screenshots of any source links.
If the post includes video, then the input includes multiple screenshots captured at intervals throughout the video.
The prompt includes clear guidance on the extent to which media in the post must match media in the source link screenshots.
For example, if a note claims that the post includes media that was taken out of context, then the sourcing must contain media that is an exact match to the post to establish the original context for the media.
The CW initially queries the Screenshot Rejector once.
If the initial query fails, then the CW issues three additional queries, all of which must pass for the proposed note to pass the Screenshot Rejector.

\textbf{Revision Rejector} economizes contributor ratings by moderating whether multiple notes can be proposed for a single post.
The Revision Rejector prompt includes the post, original note, and revised note, and guides the model to systematically identify and evaluate deltas between the notes.
Additions should introduce new claims, improve sourcing, or strengthen the argument, and removals should make the note more concise, focused, or correct.
The rejector prompt requests a continuous score, which is then scaled by a sequence matching similarity metric applied to the proposed notes with URLs removed.

\subsection{Deletion Model}

The Deletion Model identifies notes that are unlikely to obtain CRH status and proactively deletes the notes.
Deleted notes no longer show to contributors on X, helping to economize ratings, but do still count against note submission quota and the daily writing limit calculation.

The Deletion Model depends on the AI API, which exposes functionality to retrieve real-time note status and ratings for prior submissions.
Access to real-time ratings requires fulfilling the same requirement of at least 100 net CRH notes in the last 90 days that applies to the \texttt{XXL} feed.
Real-time rating information includes counts of Helpful, Somewhat Helpful, and Not Helpful ratings as well as rating tags, which capture detailed note qualities (e.g. insufficient sourcing, misses key points, etc.).
The API aggregates helpfulness and tag counts into three bins corresponding to positive, neutral, and negative values of the \textit{rater factor}, a continuous value learned from past contributor ratings that represents contributor viewpoint.

The Deletion Model predicts note status using logistic regression applied to the rating count aggregates.
The feature preparation includes both discretization and polynomial crosses, which capture interactions between features.
Since the CW applies the Deletion Model in real-time as ratings arrive, the training process avoids using the full set of ratings on historical notes.
Rather, the training data preparation extracts features up to 10 times at intervals spaced through the progression of the first 30 ratings, allowing production data to remain in-distribution as ratings accrue.

\begin{figure}
  \centering
  \includegraphics[width=\linewidth]{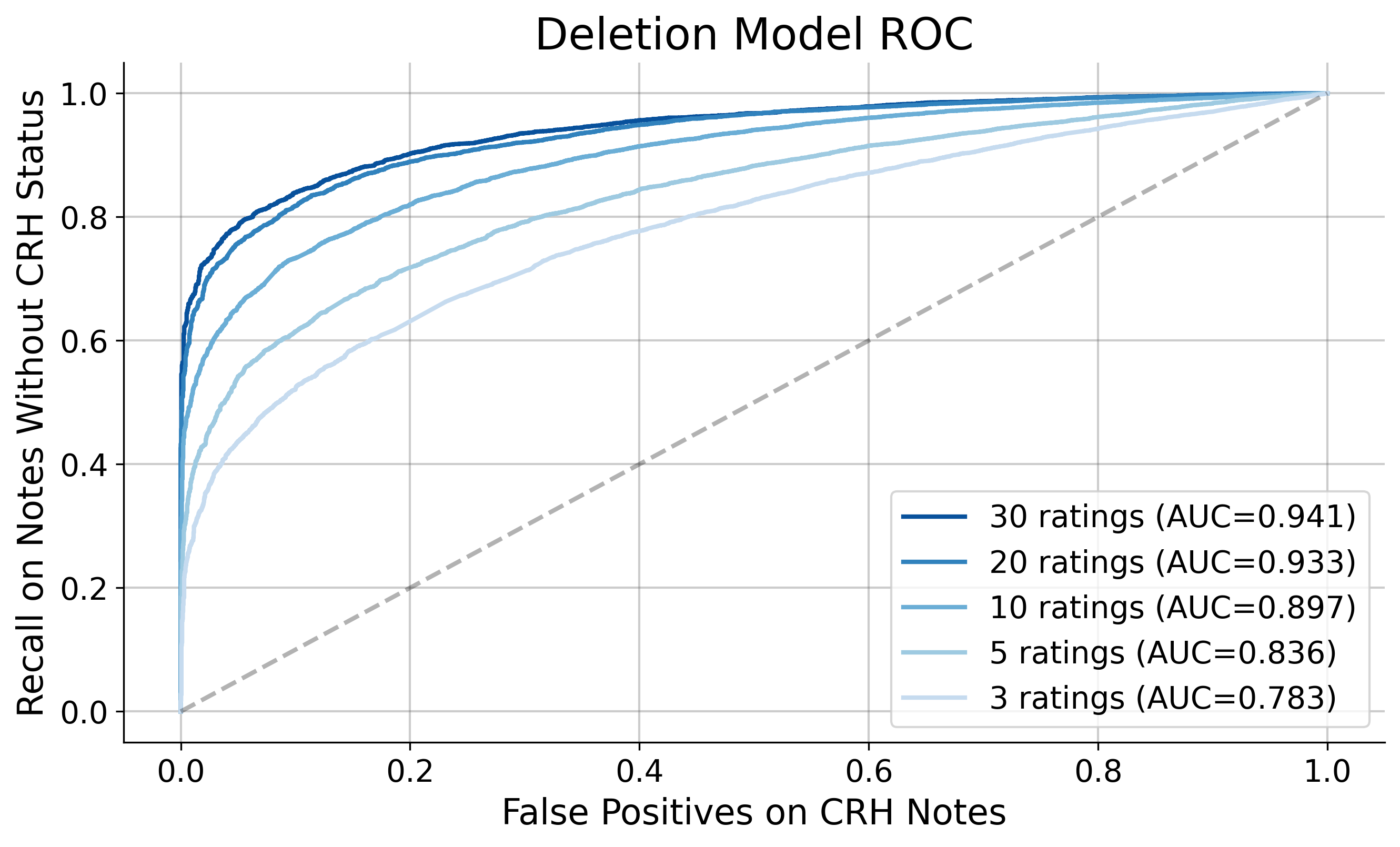}
  \caption{Deletion Model performance improves as ratings accumulate. The CW applies the Deletion Model every 60 seconds as soon as 3 ratings are available.}
  \label{fig:del-roc}
\end{figure}

Figure~\ref{fig:del-roc} presents the Deletion Model performance on evaluation data, including separate curves reflecting how many ratings were used in the prediction.
Deletion Model predictive accuracy improves as more ratings become available.
In production, the CW polls ratings every 60 seconds and applies the model as ratings arrive.
The CW includes an elevated deletion threshold, allowing deletion based on 3 or 4 ratings when the model is highly confident, as well as a standard threshold applied when 5 or more ratings are available.
The full feature extraction and training code of the Deletion Model is available in the open-source release.

\section{Evaluation}
\label{sec:eval}

This section evaluates AI writing over a 28 day period from June 2, 2026 through June 29, 2026.\footnote{Excludes \textit{collaborative notes}, a pilot feature which can generate multiple notes per post. See https://communitynotes.x.com/guide/en/contributing/collaborative-notes.}
We begin by profiling the API feeds and progression of each feed through the CW, including feed size, NPM effects, and submission outcomes.
Then, we measure the efficacy of proposed notes from the CW, including comparison with other AI and human writers.
We find the CW has been impactful and efficient, accounting for 41\% of proposed notes, 52\% of CRH notes and 55\% of CRH note views while consuming only 22\% of contributor ratings.
We also find that AI and human writers have been broadly complementary: among posts with a CRH note, 42\% have no human proposed note and 30\% have no AI proposed note.

Recall that the API includes four feed sizes, each characterized by distinct inclusion criteria.
Table~\ref{tab:feed-profile} presents the size, latency, and post coverage for each feed size.
Decreased inclusion signal requirements for larger feeds yield exponential growth, with post volume increasing approximately 3-5x across each feed size.
Correspondingly, posts also appear in larger feeds faster: the median post age decreases from 604 to 185 minutes from the \texttt{Small} to \texttt{XXL} feed, with 10\% of posts appearing in the \texttt{XXL} feed in 9 minutes or less.

Larger feeds also yield increased coverage of both posts with human written CRH notes and X platform content as a whole.
Since human contributors are able to submit notes on any post, we present the feed coverage of posts with human CRH (H-CRH) notes as a measure of how well the feed has covered posts that may receive a Community Note.
The \texttt{Small} feed contains a relatively dense concentration of misleading posts, covering 24.3\% of H-CRH notes while averaging 428 posts daily.
By comparison, the \texttt{XXL} feed is \textasciitilde57x larger, and increases coverage of H-CRH notes \textasciitilde3x to 73.4\%.
Note that all feeds skew towards content which has high visibility on X, with the \texttt{XXL} feed covering posts generating 8.1\% of X post views daily.

The CW accommodates differences in feed characteristics by customizing thresholds to each feed size.
To maintain a high CRH rate, the CW uses higher thresholds for the ClaimOpinion evaluator and Helpfulness Rejector when operating on larger feed sizes, which reflect lower levels of user demand for proposed notes.
Rejection thresholds for the \texttt{XL} and \texttt{XXL} feeds are also segmented by the NPM, with higher thresholds applied in cases where the NPM was less confident the post may receive a CRH note.

Due to differences in both post characteristics and pipeline configuration, the rate of post progression through the CW varies for each feed.
Table~\ref{tab:feed-profile} presents the progression of each feed size.
The influence of the NPM varies with feed size: the NPM is disabled for the \texttt{Small} feed, but passes only 17.1\% of posts in the \texttt{XXL} feed.
Simultaneously, the NPM effectively retains the majority of posts that receive a human CRH note, with H-CRH coverage on the \texttt{XXL} feed decreasing from 73.4\% to 55.8\% as a result of NPM filtering.
After NPM filtering, the frequency of writers generating a draft note decreases with larger feed sizes, reflecting the decreased concentration of misleading posts in larger feeds.
Rejectors ultimately mediate submission, yielding CRH rates between 7.9\% and 10.5\% for each API feed size.

The remainder of the evaluation deals with comparison between the CW and other AI and human writers.
The evaluation differentiates performance of \emph{top writers} as designated by X Community Notes, which requires achieving a lifetime hit rate of at least 4\% and net CRH of at least 10.
Top writers enjoy several special abilities, including priority in rating notifications sent to X Community Notes contributors and visibility into note requests~\cite{top_writers}.
Achieving top writer status is relatively difficult: only 3.5\% of human writers active during the evaluation period meet the top writer criteria.

\begin{figure}
  \centering
  \includegraphics[width=\linewidth]{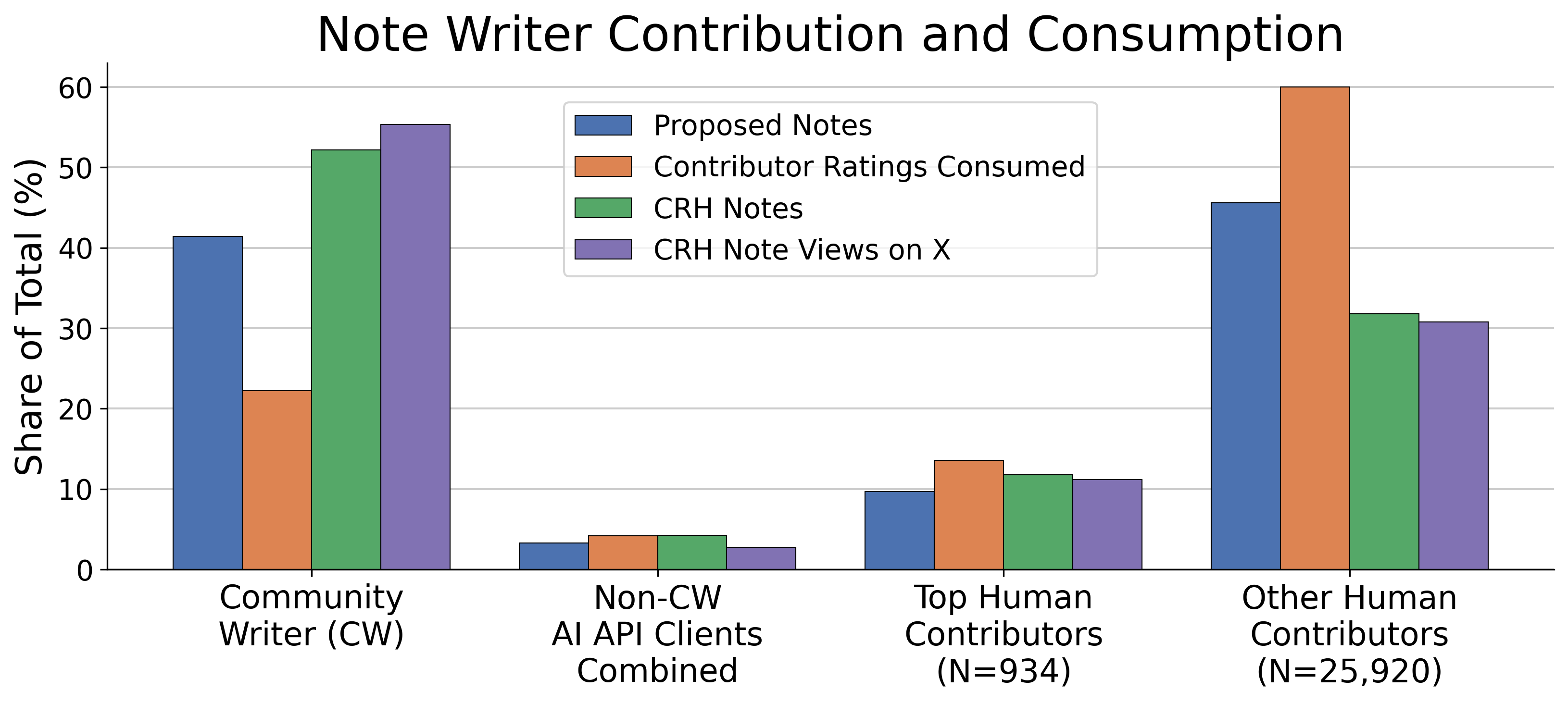}
  \caption{
    The CW maintains above-average CRH rate and impressions per CRH note, and does so while consuming below-average amounts of contributor ratings.
  }
  \label{fig:eval_overview}
\end{figure}

The CW demonstrates improvements to proposed note CRH rates, views of CRH notes on X, and contributor ratings consumed.
Figure~\ref{fig:eval_overview} details the distribution of proposed and CRH notes, contributor ratings, and CRH note views across contributor segments.
Notice that the CW accounts for 41\% of proposed notes but 52\% of CRH notes, demonstrating an above average rate of achieving CRH status.
CRH notes from the CW also account for 55\% of CRH note views on X, reflecting reduced note creation latency relative to other sources.
Simultaneously, the CW consumes only 22\% of ratings, economizing use of contributor time.

\begin{figure}
  \centering
  \includegraphics[width=\linewidth]{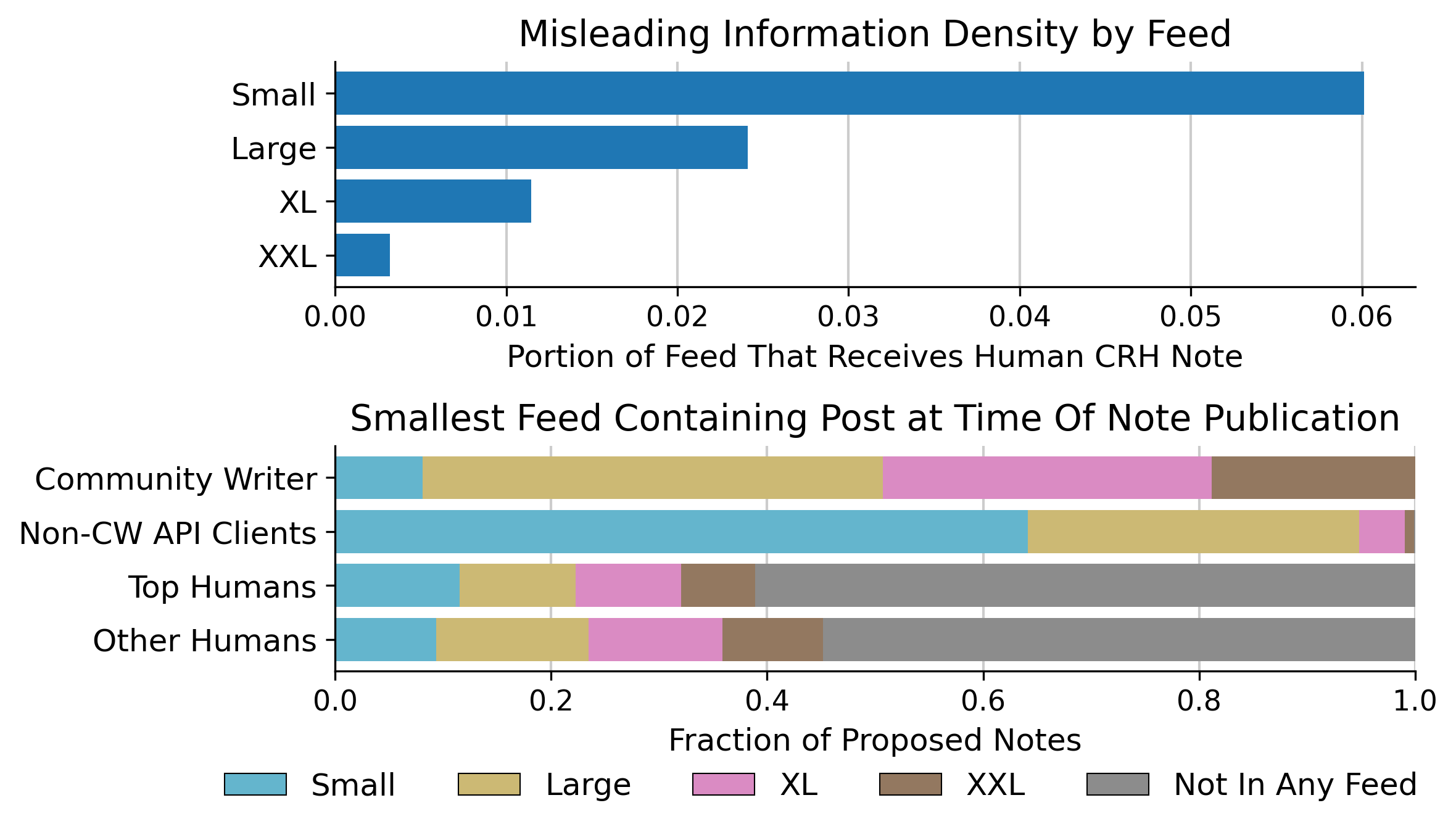}
  \caption{Decreased density in larger feeds complicates scaling, which requires progressively precise prediction of whether a draft note will achieve CRH status.}
  \label{fig:feed_overview}
\end{figure}

The NPM and rejectors allow the CW to scale by improving the CRH yield on submitted notes.
Figure~\ref{fig:feed_overview} presents the density of posts with human CRH notes in each feed and the distribution of feed adoption by different writers.
AI writers other than the CW submit 64\% of notes from the \texttt{Small} feed and 31\% of notes from the \texttt{Large} feed, which have a higher density of posts with human CRH notes than the XL or XXL feeds.
The NPM and rejection mechanisms accommodate the class imbalance of the XL and XXL feeds, allowing the CW to process larger feeds without exhausting note writing quota.
Figure~\ref{fig:feed_overview} provides comparisons to human writers as a reference; humans are free to write on any post at any time.

\begin{figure}
  \centering
  \includegraphics[width=\linewidth]{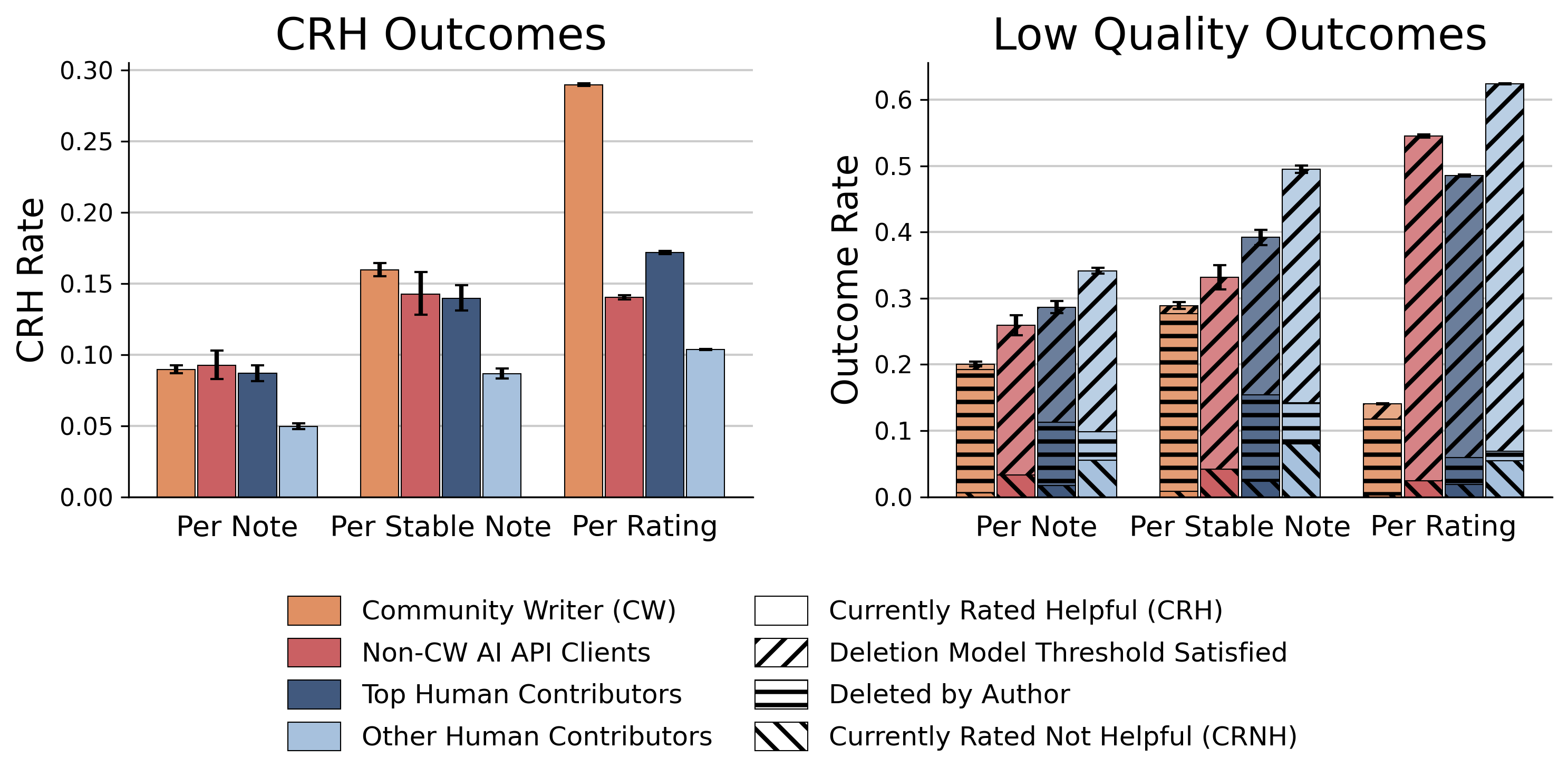}
  \caption{Stable notes include any note that is CRH, CRNH, deleted or meets a rating minimum.  Conditioned on stability, the CW outperforms on both CRH and low quality outcomes.}
  \label{fig:quality}
\end{figure}

Despite processing feeds with a lower density of misleading posts, CW proposed note quality compares well with other AI writers and humans.
The \textit{Per Note} portions of Figure~\ref{fig:quality} present the rates of CRH and \textit{low quality} outcomes over all proposed notes.\footnote{Error bars represent 95\% Wilson confidence interval.}
We define low quality outcomes as any note which is either CRNH or deleted by the author.
The Deletion Model results in higher deletion rates and lower CRNH rates since many notes are deleted before reaching CRNH status.
For comparison, Figure~\ref{fig:quality} simulates applying the Deletion Model to all writers using historical data.

Variations in the visibility of posts complicate comparison of note status outcomes, which effectively require 5 and 10 ratings to reach CRNH and CRH status respectively.
Figure~\ref{fig:quality} presents outcomes for \textit{stable notes}, defined as any note that is CRH, CRNH, deleted by the author or has at least 5 ratings (low quality notes) or 10 ratings (CRH notes).
Measurement over stable notes increases the rates of CRH and low quality outcomes across all writers.
The CW maintains the highest CRH rate\footnote{Two-sided z-test comparing with Non-CW AI API clients indicated significant difference, with $z$=2.082 and $p$=.037.} despite having more deleted notes, many of which are below rating thresholds.

Since note status depends on contributor willingness to supply ratings, Figure~\ref{fig:quality} also presents the rate of status outcomes \textit{per rating}.
Measuring outcomes per rating captures the experience of contributors, reflecting the likelihood that the effort to make a rating yields a CRH note.
The CW offers a 29\% likelihood of ratings leading to a CRH note, compared to 17\% for notes written by top writers.

\begin{figure}
  \centering
  \includegraphics[width=\linewidth]{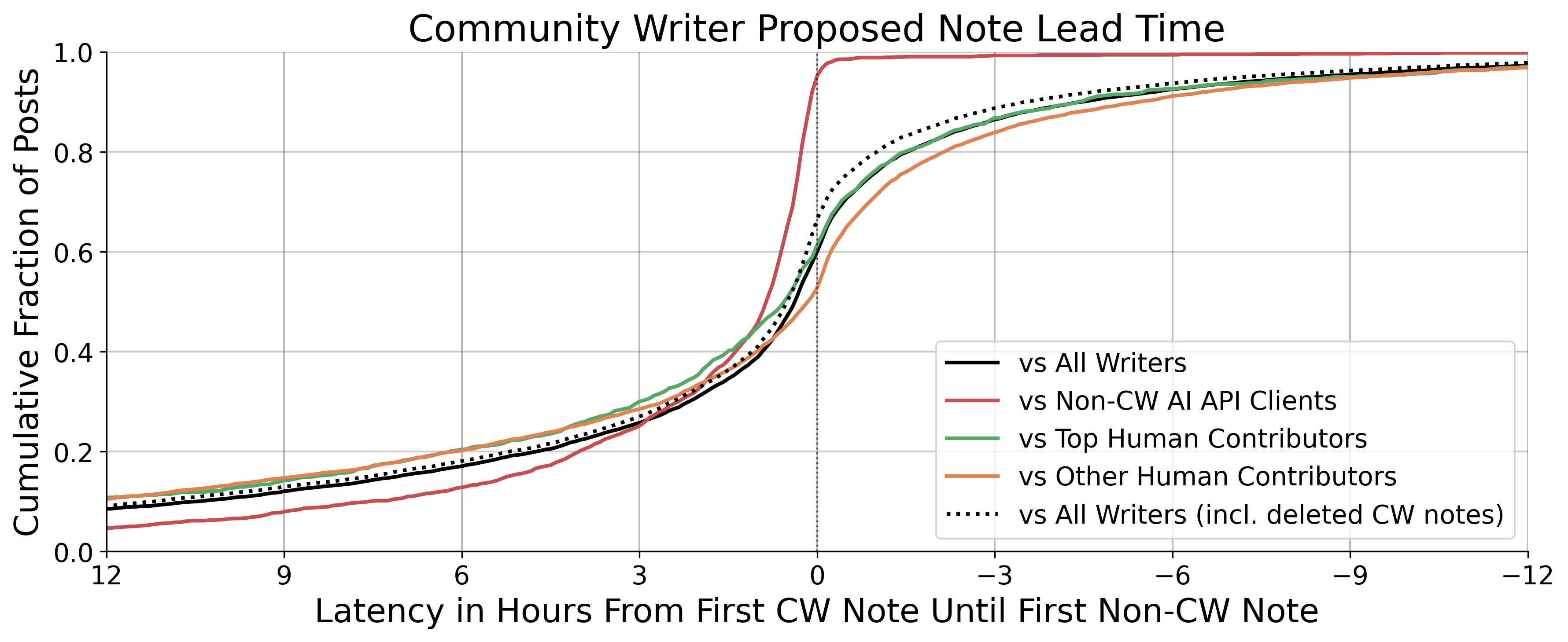}
  \caption{Compared to other writers, the CW submits the first proposed, non-deleted note on 60\% of posts.}
  \label{fig:latency}
\end{figure}

Operating on larger feeds improves latency, as decreased user signal requirements allow posts to enter feeds sooner.
Figure~\ref{fig:latency} presents the amount of \emph{lead time} by which CW proposed notes precede proposed notes from other sources, including the effect of CW deletions.
By design, the plot can only include posts where there is a proposed note from both the CW \emph{and} another writer.
Among non-deleted notes, the CW publishes the first proposed note on 60\% of posts and the first AI proposed note on 95\% of posts, leading to CRH status faster and increasing views of CRH notes.

\begin{figure}
  \centering
  \includegraphics[width=\linewidth]{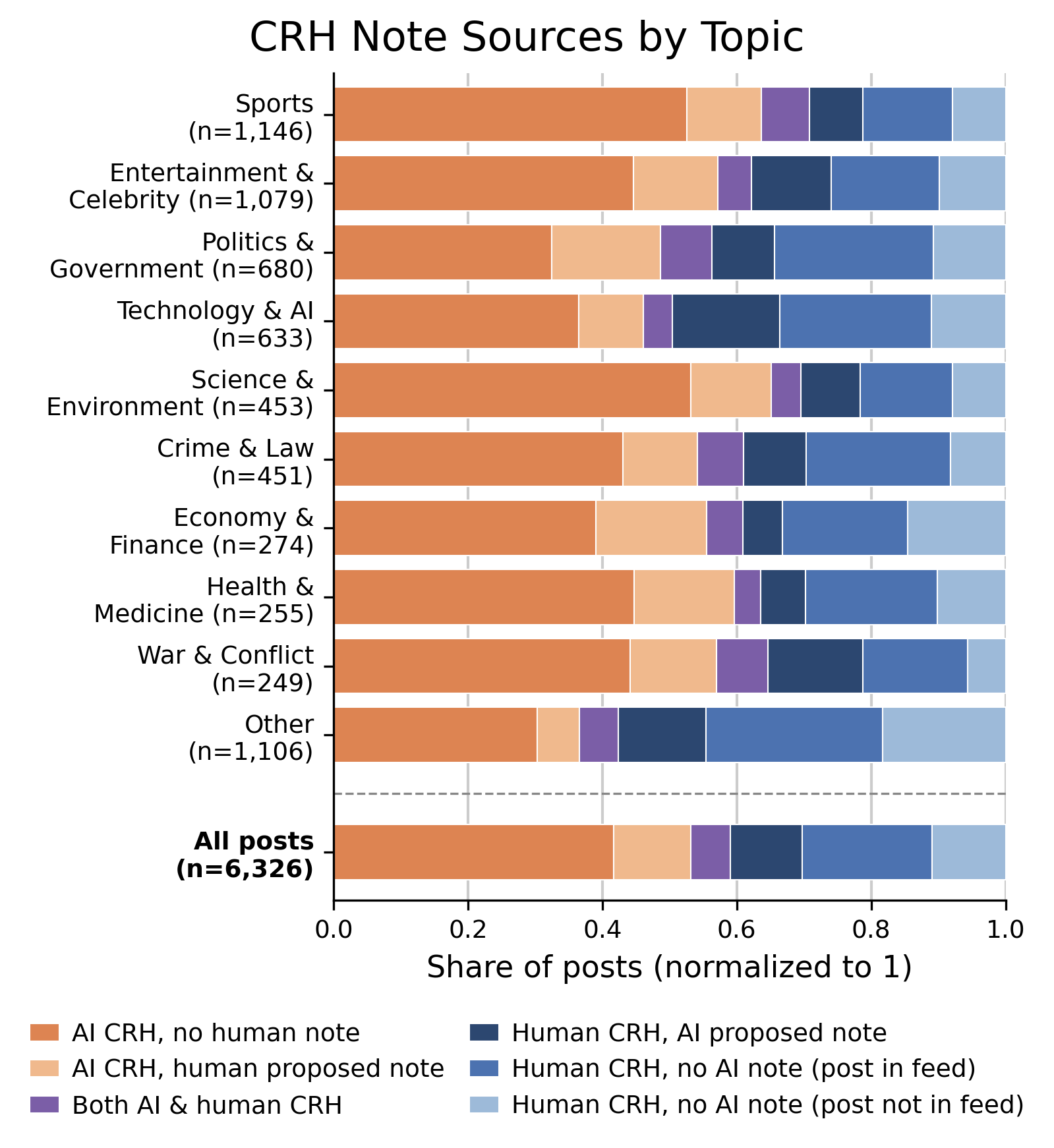}
  \caption{AI writers extend coverage, with most AI CRH notes occurring on posts without a human proposed note.}
  \label{fig:topics}
\end{figure}

AI writers provide broad impact across topics in Community Notes, most often complementing the work of human writers.
Figure~\ref{fig:topics} presents the distribution of CRH note sources across 10 topics, with post topics identified and assigned by Grok 4.20.
Most AI CRH notes occur on posts where there is no proposed human note, indicating that human contributors saw the AI note, found it helpful, and decided not to propose an alternative.
Simultaneously, human writing remains a critical contribution, fundamentally extending the coverage of Community Notes: 41\% of posts with CRH notes have no AI written CRH note, and 30\% have no AI proposed note entirely, inclusive of deleted CW notes.

\section{Discussion and Limitations}
\label{sec:discussion}

\textbf{Community Guidance and Authority}
AI writers add note production beyond humans, while remaining subject to human authority.
The community guides the CW in particular through API feeds, model training, and model inputs, focusing notes according to community taste and demand.
AI writers have been effective, supporting 86.0\% growth in CRH notes shown broadly on X.\footnote{Reflects all AI and human writers combined, with 28-day trailing averages as of September 1, 2025 and September 20, 2026 and note status as of September 21, 2026.}
Human CRH notes continue, often thriving on topics that are rapidly evolving or require nuanced awareness.
All notes remain subject to human raters, whose ratings demonstrate the utility of AI generated notes and the ongoing value of human note writing.

\textbf{Contributor Experience}
Scaling AI note writing requires ratings from people who volunteer their time and attention to shape Community Notes.
Recognizing that rater attention is a critical resource, the Notable Post Model and Deletion Model yield direct savings by focusing ratings on notes most likely to become CRH.
Beyond immediate gains, rating a note that becomes CRH and displays broadly on X is satisfying to contributors, demonstrating their time and attention were productive and well used, potentially yielding improved retention and engagement among Community Notes volunteers.

\textbf{Limitations}
While this work presents and details AI note writing at scale, comparison between writing techniques is limited.
We do not control for differences in the operational cost or post feed sizes of AI writers and compare the Community Writer to all other AI writers \textit{in aggregate}.
Segmenting AI writers is difficult, as AI writers may change over time, operate multiple accounts, and combine multiple writing behaviors within a single account.
We are pursuing future work which will create a direct, principled forum for comparing AI writing techniques to advance open research while producing the best notes for contributors and people on X. 

\section{Conclusion}
\label{sec:conclusion}
This work presents both the Community Notes AI API and Community Writer, which combine to offer an effective, open approach to increase the coverage of Community Notes on X.
We demonstrate that AI note writing complements notes from human contributors, offering improvements to speed, rating efficiency, and the yield of Currently Rated Helpful notes shown broadly on X.
While AI offers strengths, human contributors remain the core and ultimate authority of Community Notes, continuing to govern the display of both AI and human proposed notes.

\bibliographystyle{ACM-Reference-Format}
\bibliography{references}

\appendix

\section{API Quota Algorithm}
\label{sec:algo}

\begin{algorithm}[H]
\caption{Progressive writing quota}
\label{alg:quota}
\begin{algorithmic}[1]
\Statex $NH_5,NH_{10} \gets$ Num. CRNH in the last 5, 10 CRH/CRNH notes
\Statex $HR_{20},HR_{100} \gets$ (CRH-CRNH)/Total over last 20, 100 notes
\Statex $HR_{14d} \gets$ (CRH-CRNH)/Total over last 14 days among notes with at least 10 ratings, CRH or CRNH status
\Statex $DN \gets$ Average daily notes written in last 30 days
\Statex $T \gets$ Total notes ever submitted
\Statex $WL \gets$ Daily writing limit
\Statex
\If{$NH_{10} \ge 8$} \Comment{Quality regression killswitch}
  \State $WL \gets 2$
\ElsIf{$NH_5 \ge 3$}
  \State $WL \gets 5$
\ElsIf{$T < 20$} \Comment{New writer}
  \State $WL \gets 10$
\Else \Comment{Quota ramp-up}
  \State  $HR \gets \max(HR_{14d}, HR_{100})$
  \If{$HR < 0.05$}
    \State $WL \gets 300 \cdot \max(HR_{20}, HR)$
  \ElsIf{$HR < 0.10$}
    \State $WL \gets 15 + 700\,(HR - 0.05)$
  \ElsIf{$HR < 0.15$}
    \State $WL \gets 50 + 3000\,(HR - 0.10)$
  \ElsIf{$HR < 0.20$}
    \State $WL \gets 200 + 6000\,(HR - 0.15)$
  \Else
    \State $WL \gets 500$
  \EndIf
  \State $WL \gets \max\!\big(5,\ \lfloor \min(5\,DN,\ WL) \rfloor\big)$ \Comment{Limit rate of change}
\EndIf
\end{algorithmic}
\end{algorithm}

\section{Ratings per Status Outcome}
\label{sec:bonus_plots}

\begin{figure}[h!]
  \centering
  \includegraphics[width=\linewidth]{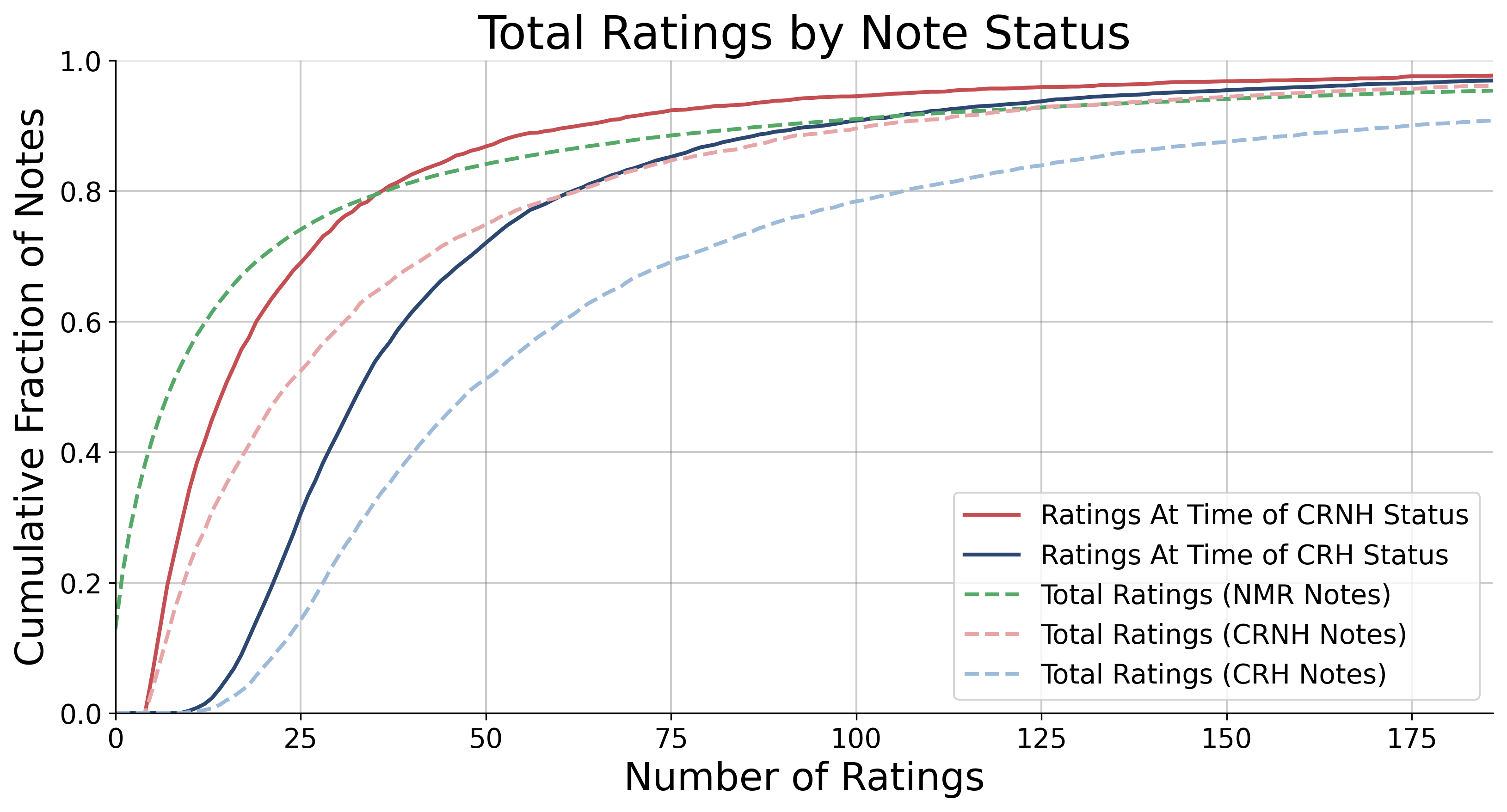}
  \caption{54\% of NMR notes receive fewer than 10 ratings.}
  \label{fig:min_ratings}
\end{figure}

While the AI API offers a scalable approach to increasing the supply of notes, contributor ratings remain the sole mechanism that allows notes to show broadly on X.
In effect, notes generally require at least 10 ratings to achieve CRH status and at least 5 ratings to achieve CRNH status.
During the evaluation period from June 2-29, 2026, 54\% of NMR notes receive fewer than 10 ratings, which effectively prevents the notes from reaching CRH status.
Low rating counts reflect the free and voluntary nature of contributing to Community Notes: contributors are free to propose notes on low visibility posts and free to decline to rate notes that they don't find helpful.
Figure~\ref{fig:min_ratings} presents the distribution of rating counts for CRH, CRNH, and NMR notes, excluding any notes that were deleted, along with curves detailing the number of ratings when a note first achieved CRH or CRNH status.

\end{document}